\documentclass[utf8]{frontiersFPHY} 

\usepackage{url,hyperref,lineno,microtype,subcaption}
\usepackage[onehalfspacing]{setspace}
\usepackage{soul}

\def\keyFont{\fontsize{8}{11}\helveticabold }
\def\firstAuthorLast{Held {et~al.}} 
\def\Authors{Karsten Held\,$^{1,*}$,  Liang Si$^{1,2,*}$, Paul Worm$^{1}$, Oleg Janson$^{3}$,  Ryotaro Arita$^{4,5}$,  Zhicheng Zhong$^{6}$, Jan M. Tomczak$^{1}$, {Motoharu Kitatani}$^{1,4,*}$}
\def\Address{$^1$ Institute for Solid State Physics, Vienna University of Technology, 1040 Vienna, Austria \\  $^2$ School of Physics, Northwest University, Xi'an 710069, China
 \\   $^3$ {Institute for Theoretical Solid State Physics, Leibniz IFW Dresden, 01069} Dresden, Germany \\
  $^4$  RIKEN Center for Emergent Matter Sciences (CEMS), Wako, Saitama, 351-0198, Japan \\ 
  $^5$  Department of Applied Physics, The University of Tokyo, Hongo, Tokyo, 113-8656, Japan \\
  $^6$ Key Laboratory of Magnetic Materials and Devices \& Zhejiang Province Key Laboratory of Magnetic Materials and Application Technology, Ningbo Institute of Materials Technology and Engineering (NIMTE), Chinese Academy of Sciences, Ningbo 315201, China }
\def\corrAuthor{{Karsten Held, Liang Si, or Motoharu Kitatani}}

\def\corrEmail{{held@ifp.tuwien.ac.at, liang.si@ifp.tuwien.ac.at, motoharu.kitatani@riken.jp}}

\begin{document}
\onecolumn
\firstpage{1}

\title[Nickelate superconductors calculated by D$\Gamma$A]{Phase diagram of nickelate superconductors calculated by dynamical
vertex approximation} 

\author[\firstAuthorLast ]{\Authors} 
\address{} 
\correspondance{} 

\extraAuth{}

\maketitle

\begin{abstract}

\section{}
We review the electronic structure  of nickelate superconductors with and without effects of
{electronic}
correlations. As a minimal
model we identify the one-band Hubbard model for the Ni 3$d_{x^2-y^2}$
orbital plus a pocket around the  $A$-momentum. The latter however merely
acts as a decoupled electron reservoir. This reservoir makes a careful
translation from {nominal} Sr-doping to the doping of the one-band Hubbard model
mandatory. Our dynamical mean-field theory calculations, in part already
supported by experiment, indicate that the $\Gamma$ pocket, {Nd}
4$f$ orbitals, oxygen 2$p$ and {the} other Ni 3$d$ orbitals are not relevant in
the superconducting doping regime. The physics is completely different if topotactic hydrogen is present or the oxygen reduction is
incomplete. 
{Then,}
a two-band physics {hosted by} the Ni  3$d_{x^2-y^2}$ and 3$d_{3z^2-r^2}$
orbitals {emerges}.
Based on our minimal modeling we calculated the superconducting $T_c$
vs.\ Sr-doping $x$ phase diagram prior to experiment using the dynamical
vertex approximation. For such a notoriously difficult to determine
quantity as $T_c$, the agreement with experiment is astonishingly good.  The prediction that $T_c$ is enhanced with pressure or compressive strain, has been confirmed experimentally as well.
This supports that the one-band Hubbard model plus {an} electron
reservoir is the appropriate minimal model.

\tiny
 \keyFont{ \section{Keywords:} nickelate superconductivity, dynamical mean-field theory, quantum field theory, computational materials science, density functional theory} 
\end{abstract}

\section{Introduction}
Twenty years ago,  Anisimov, Bukhvalov, and Rice~\cite{Anisimov1999} suggested high-temperature ($T_c$) superconductivity in nickelates based {on} {material calculations} {that showed} apparent similarities to cuprates. Subsequent calculations~\cite{Chaloupka2008,PhysRevLett.103.016401,Hansmann2010b} demonstrated the potential to further engineer the nickelate Fermi surface through heterostructuring.  Two years ago
Li, Hwang and coworkers~\cite{li2019superconductivity} discovered superconductivity in Sr-doped NdNiO$_2$ films grown on a SrTiO$_3$ substrate
and protected by  a  SrTiO$_3$ capping layer.
These novel Sr{$_x$}Nd$_{1-x}$NiO$_2$ films
are isostructural  and  formally isoelectric to the arguably simplest, but certainly not best superconducting cuprate:  infinite layer CaCuO$_2$~\cite{siegrist1988parent,Balestrino2002,Orgiani2007,DiCastro2015}.

\begin{figure*}[tb]
 \begin{minipage}{.64\textwidth} \hfill
   \includegraphics[width=\textwidth]{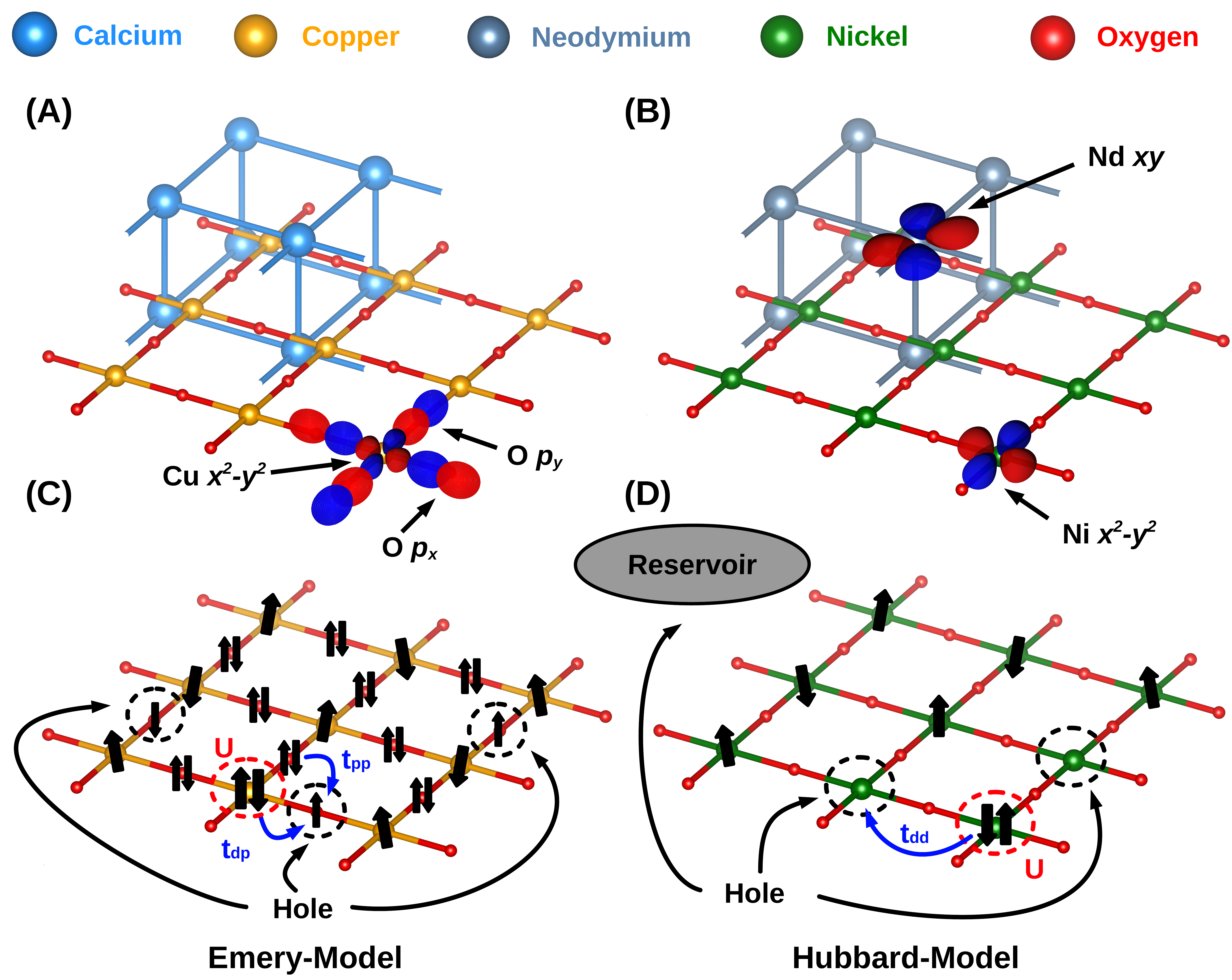}\hfill
 \end{minipage}\hfill
 \begin{minipage}{.35\textwidth}
 \vspace{.2cm}
   \caption{\label{Fig:Model}Crystal lattice and {most important orbitals} for (A)  cuprates  and  (B) nickelates. (C) For cuprates the arguably  simplest model is the {\em Emery model} with {a} half-filled copper 3$d_{x^2-y^2}$ band and holes in the oxygen 2$p$ orbitals that can hop to other oxygen ($t_{pp}$) and copper sites ($t_{dp}$) where double occupations are suppressed by the interaction $U$. (D) For nickelates  we have a  Ni-3$d_{x^2-y^2}$-band  {\em Hubbard model} which however only accommodates part of the holes induced by  Sr-doping. The others go to the $A$ pocket {stemming} from {the} Nd 5$d_{xy}$ {band} and acting as a decoupled {\em reservoir}. }
   \end{minipage}
\end{figure*}

However, the devil is in the details, and here cuprates and nickelates differ.
For revealing such material-specific differences,
band{-}structure calculations based on density functional theory (DFT)
are the method of choice. They serve as a starting point for understanding the
electronic structure and subsequently the phase diagram of nickelate superconductors. Following the experimental discovery of nickelate superconductivity, and even before that, numerous  DFT studies have been published~\cite{Pavarini2001,Lee2004,Botana2019,Hirofumi2019,Motoaki2019,hu2019twoband,Wu2019,Nomura2019,Zhang2019,Jiang2019,Werner2019}. Based on these DFT calculations, various models for the low-energy electronic structure for nickelates and the observed superconductivity have been proposed. Besides the cuprate-like
Ni  3$d_{x^2-y^2}$ band, DFT shows an $A$ and a $\Gamma$ pocket which originate from Nd 5$d_{xy}$ and 5$d_{3z^2-r^2}$ bands, but with major Ni 3$d$ admixture in the region of the pocket.  The importance of the Ni 3$d_{3z^2-r^2}$ orbital has been suggested in some studies~{\cite{Lechermann2019,Lechermann2020,Petocchi2020,Adhikary2020}} and that of  the Nd-$4f$ orbitals in others~{\cite{Zhang2019,Subhadeep2020}}. Further there is the question regarding the relevance of the oxygen 2$p$ orbitals. For cuprates these are, besides the Cu  3$d_{x^2-y^2}$ orbitals, the most relevant.
 Indeed, cuprates  are generally believed to be  charge{-}transfer insulators~\cite{Zaanen1985}.
This leads  to the three-orbital {\em Emery model}~\cite{Emery1987} visualized in Fig.~\ref{Fig:Model}A,C as the minimal model for cuprates.
The much more frequently investigated Hubbard model~\cite{Hubbard1963,Kanamori63,Gutzwiller63} may, in the case of cuprates,  only  be considered as an effective Hamiltonian mimicking the physics of the Zhang-Rice singlet~\cite{Zhang1988}.

At first glance, nickelates appear to be much more complicated with more relevant orbitals than in the case of the cuprates. In this paper, we review the
electronic structure of nickelates in comparison to that of cuprates, and 
the arguments for a simpler description of nickelate superconductors: namely a {\em Hubbard model} for the Ni 3$d_{x^2-y^2}$ band plus a largely decoupled reservoir corresponding to the  $A$ pocket. {This $A$ pocket is part of the Nd $5d_{xy}$ band which has however a major admixture of Ni $3d_{xz/yz}$ and O $2p_z$ states around the momentum $A$.} This leaves us with  Fig.~\ref{Fig:Model}B,D as the arguably simplest model for nickelates~\cite{Karp2020,Kitatani2020}. This (our) perspective  is still controversially discussed in the literature. However, as we will point out below, a number of experimental observations already support this perspective against
some of the early  suggestions that other orbitals are relevant. Certainly, other perspectives will be taken in other articles of this series on ``Advances in Superconducting Infinite-Layer and Related Nickelates''.  The simple picture of a one-band Hubbard model, whose doping needs to be carefully calculated since part of the holes in Sr-doped  Sr$_x$Nd$_{1-x}$NdO$_2$ go to the $A$ pocket, allowed us~\cite{Kitatani2020} to calculate 
$T_c$, see Fig.~\ref{FigTc} below, at a time when only the $T_c$ for a single doping
$x=20\%$ was experimentally available. To this end, state-of-the-art dynamical vertex approximation (D$\Gamma$A)~\cite{Toschi2007,Katanin2009,RMPVertex}, a Feynman diagrammatic extension of dynamical mean-field theory (DMFT)~\cite{Metzner1989,Jarrell1992,Georges1992,Georges1996} has been used. For such a  notoriously difficult to calculate physical quantity as $T_c$, the agreement of the single-orbital Hubbard model calculation with subsequent experiments~\cite{Li2020,zeng2020} is astonishingly good. This further supports the modelling by a single-orbital Hubbard model which thus should  serve at the very least as a  good approximation or {a} starting point.

The outline of this article is as follows: In Section \ref{Sec:comp} we first compare the electronic structure of nickelates to that of cuprates, starting from DFT but also discussing effects of electronic correlations as described, e.g., by DMFT.
Subsequently, we argue in  Section \ref{Sec:other}, orbital-by-orbital, that the other orbitals besides the  Ni 3$d_{x^2-y^2}$ and  the $A$ pocket are, from our perspective, not relevant.  This leaves us with the one-3$d_{x^2-y^2}$-band Hubbard model plus an electron reservoir representing the $A$ pocket of Fig.~\ref{Fig:Model}B,D, which is discussed in  Section \ref{Sec:fill} including the translation of Sr-doping to the filling in the Hubbard model and the reservoir.
 In  Section \ref{Sec:DGA}, we discuss the effect of non-local correlations as described in  D$\Gamma$A and the calculated superconducting phase diagram.   Section \ref{Sec:topH} shows that topotactic hydrogen, {which}
is difficult to detect in experiment{,} completely overhauls the electronic structure  and the prevalence {of} superconductivity. Finally,   Section \ref{Sec:conclusio} summarizes the article.

\section{Electronic structure: Nickelates vs.\ cuprates}
\label{Sec:comp}

\begin{figure*}[tb]
 \begin{minipage}{.5205\textwidth} 
   \includegraphics[width=1.\textwidth]{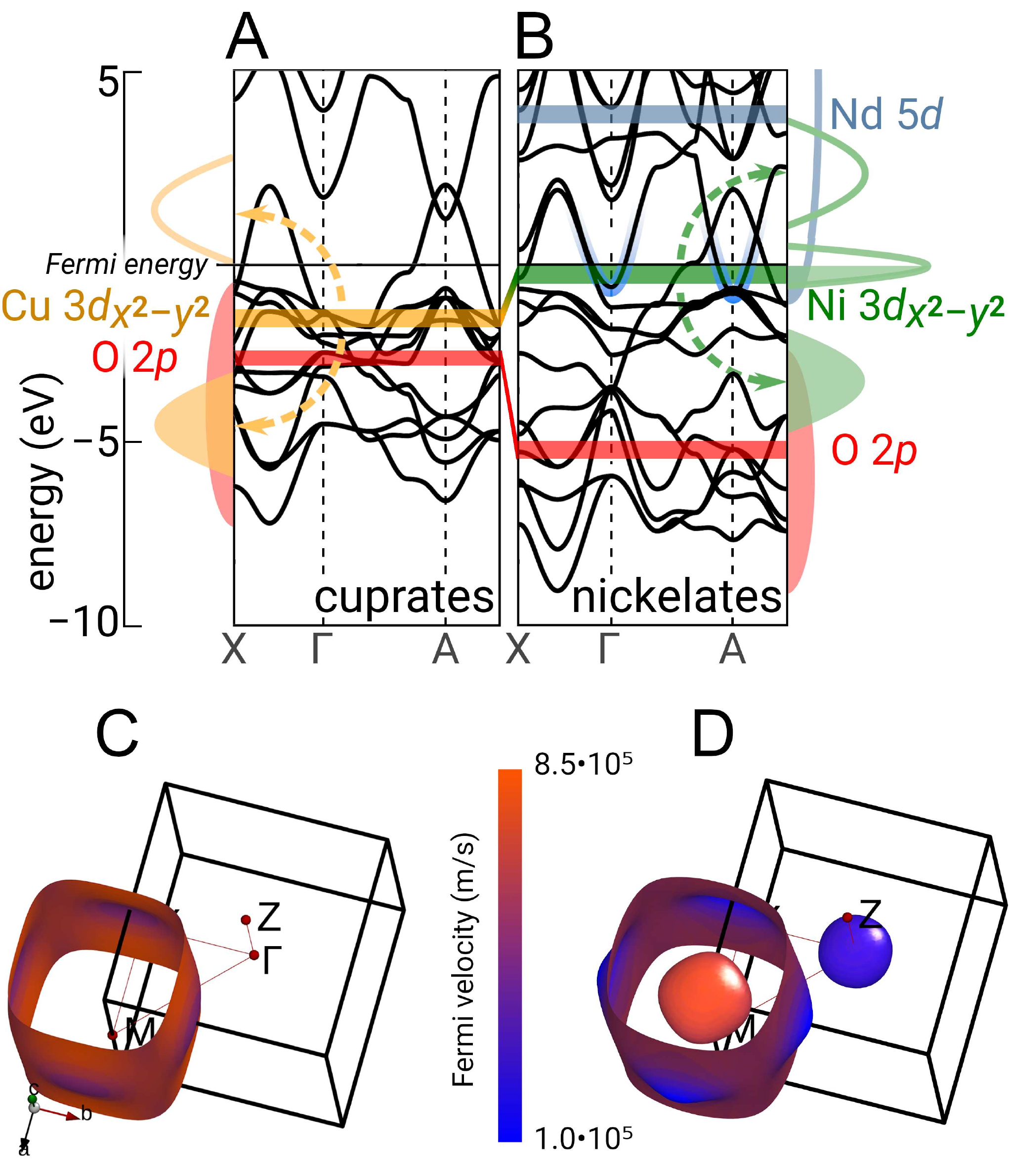}
 \end{minipage}\hfill
 \begin{minipage}{.444\textwidth}
   \caption{\label{Fig:DFTbands} \normalsize Electronic structures of CaCuO$_2$ (A) and NdNiO$_2$ (B), exemplifying superconducting cuprates and nickelates. Top: The bars indicate the center of energy for the most important DFT bands. The dashed arrows indicate the  correlation-induced splitting of the   Cu or Ni 3$d_{x^2-y^2}$ band into Hubbard bands, leading to the  (schematic)  spectral function  at the side. Cuprates are charge-transfer insulators with a gap between   the  O 2$p$  and the upper  Cu  3$d_{x^2-y^2}$ Hubbard band,  {as} the lower Cu  3$d_{x^2-y^2}$ Hubbard band is below the  O 2$p$ band. For nickelates,
     bands  with a large Nd(La) 5$d$ contribution  cross  the Fermi energy at $\Gamma$ and $A$, self-doping the  Ni 3$d_{x^2-y^2}$ band away from half-filling. C and D: Corresponding DFT Fermi surface. For cuprates DFT shows a single  Cu  3$d_{x^2-y^2}$ hole-like Fermi surface; for nickelates there is a similar, slightly more warped  Ni  3$d_{x^2-y^2}$ Fermi surface and additional small pockets around the $\Gamma$ and $A$ momentum.}
   \end{minipage}
\end{figure*}

Let us start by looking into the electronic structure in more detail and start with the DFT results. On a technical note, the calculations presented have been done using the  \textsc{wien2k}~\cite{blaha2001wien2k,Schwarz2002},  \textsc{VASP}~\cite{PhysRevB.48.13115}, and \textsc{FPLO}~\cite{FPLO} program packages, with the  PBE~\cite{PhysRevLett.77.3865}  version of the generalized gradient approximation (GGA). For further details see the original work~\cite{Kitatani2020}. Fig.~\ref{Fig:DFTbands} compares the bandstructure of the two simple materials: CaCuO$_2$ and NdNiO$_2$.  Here, we restrict ourselves to only the Brillouin zone path along the  most relevant momenta for these compounds: $\Gamma$ (0,0,0), $X$ ($\pi$,0,0), and $A$ ($\pi$,$\pi$,$\pi$). In DFT both the cuprate and nickelate parent compounds  are metals with a prominent Cu or Ni  3$d_{x^2-y^2}$  band crossing the Fermi energy. In other aspects both materials differ (for a review cf.~\cite{Nomura2021}):  In the case of {\em cuprates}, the {oxygen bands} are much closer to the Fermi energy. Hence, if electronic correlations split the {DFT} bands into two Hubbard bands as indicated in Fig.~\ref{Fig:DFTbands} by the arrows and the spectral function in the left side panel, we get a charge{-}transfer insulator~\cite{Zaanen1985}. For this charge{-}transfer insulator, the oxygen 2$p$ orbitals are the first orbitals below the Fermi level ($E_F$)  and receive  the holes that are induced by doping. The  Cu  3$d_{x^2-y^2}$   lower Hubbard band is below these oxygen orbitals, and the   Cu  3$d_{x^2-y^2}$  upper Hubbard band is above the Fermi level.
Let us note that we here refer to  oxygen 2$p$ orbitals and Cu  3$d_{x^2-y^2}$ orbitals even though the hybridization between both is very strong. 
{Indeed,}
the two sets of orbitals strongly mix in the resulting  effective DFT bands of Fig.~\ref{Fig:DFTbands}.

Because cuprates are  charge{-}transfer insulators, the one-band  Hubbard model can only  be considered as an effective Hamiltonian mimicking the Zhang-Rice singlet~\cite{Zhang1988}.  As already pointed out in the Introduction, more appropriate is the  {\em Emery model} of  Fig.~\ref{Fig:Model}. The correlation-induced splitting into the Hubbard bands~\cite{Georges1992} as well as the Zhang-Rice singlet~\cite{Hansmann2014} can be described already by {DMFT}~\cite{Metzner1989,Georges1992,Georges1996}. Two-dimensional spin-fluctuations and superconductivity, however, cannot. For describing such physics, non-local correlations beyond DMFT are needed.

For the {nickelates}, the oxygen bands are at a much lower energy. Hence, as indicated in the right side panel of Fig.~\ref{Fig:DFTbands}, the lower Ni 3$d_{x^2-y^2}$ Hubbard band can be expected to be closer to the Fermi energy than the oxygen $p$ orbitals~\cite{Karp2020,Kitatani2020}. Consequently, undoped nickelates would be Mott-Hubbard insulators if it was not for  two additional bands that cross $E_F$  around  the $\Gamma$- and $A$-momentum. These form electron pockets as visualized in Fig.~\ref{Fig:DFTbands} (bottom right)  and self-dope the  Ni 3$d_{x^2-y^2}$ band away from half-filling. As the  3$d_{x^2-y^2}$ is doped, it develops, even when the Coulomb interaction is large, a quasiparticle peak at the Fermi energy as displayed  in the right side panel of Fig.~\ref{Fig:DFTbands}.

\section{Irrelevance of various orbitals}
\label{Sec:other}
Next, we turn to various orbitals that may appear  relevant at first glance but turn out to be
irrelevant for the low energy physics {when taking electronic correlations properly into account.}  
{To account for the latter, we use} {DFT+DMFT~\cite{Anisimov1997,Lichtenstein1998,Held2006,Kotliar2006,Held2007} which is
the state of the art for calculating correlated materials.}

\paragraph*{Oxygen orbitals.}
For nickelates the oxygen 2$p$ orbitals are approximately 3$\,$eV lower in energy than  in cuprates within DFT. Hence, some DFT+DMFT calculations did not include these from the beginning~\cite{Karp2020,Kitatani2020}, and those that did~\cite{Lechermann2019} also found the oxygen 2$p$ orbitals at {a} lower energy than the  lower  Ni 3$d_{x^2-y^2}$ Hubbard band. Hence, while there is still some hybridization and mixing between the O 2$p$ states and  the Ni 3$d_{x^2-y^2}$ states, a  projection onto a low-energy set of orbitals without  oxygen appears possible.

\paragraph*{Ni 3$d_{3z^2-r^2}$ and $t_{2g}$ orbitals.}
Instead of the oxygen 2$p$ orbitals, the DFT calculation in 
Fig.~\ref{Fig:DFTbands} and elsewhere~\cite{Botana2019,Hirofumi2019,Motoaki2019,Nomura2021} show other Ni 3$d$ orbitals closely below the  Ni 3$d_{x^2-y^2}$ band. In fact, these other 3$d$ orbitals are  somewhat closer to the Fermi level than in the case of cuprates.  Electronic correlations can strongly modify the DFT band structure. In particular, the Hund's exchange $J$ tends to drive the system  toward a more equal occupation of different orbitals, especially if there is more than one hole (more than one unpaired electron) in the Ni 3$d$ orbitals. This is not only because a larger local spin is made  possible, but also because the inter-orbital Coulomb interaction $U'$ between two electrons in two different  orbitals  is smaller  than the intra-orbital Coulomb interaction $U=U'+2J$ for two electrons in the same orbital. This tendency is countered by   the crystal field splitting (local DFT potentials) which puts the 3$d_{x^2-y^2}$ orbital above the Ni 3$d_{3z^2-r^2}$ orbital and the other ($t_{2g}$)  Ni 3$d$ orbitals because of the absence of apical O atoms in NiO$_4$ squares .

\begin{figure*}[tb]
\includegraphics[width=18cm]{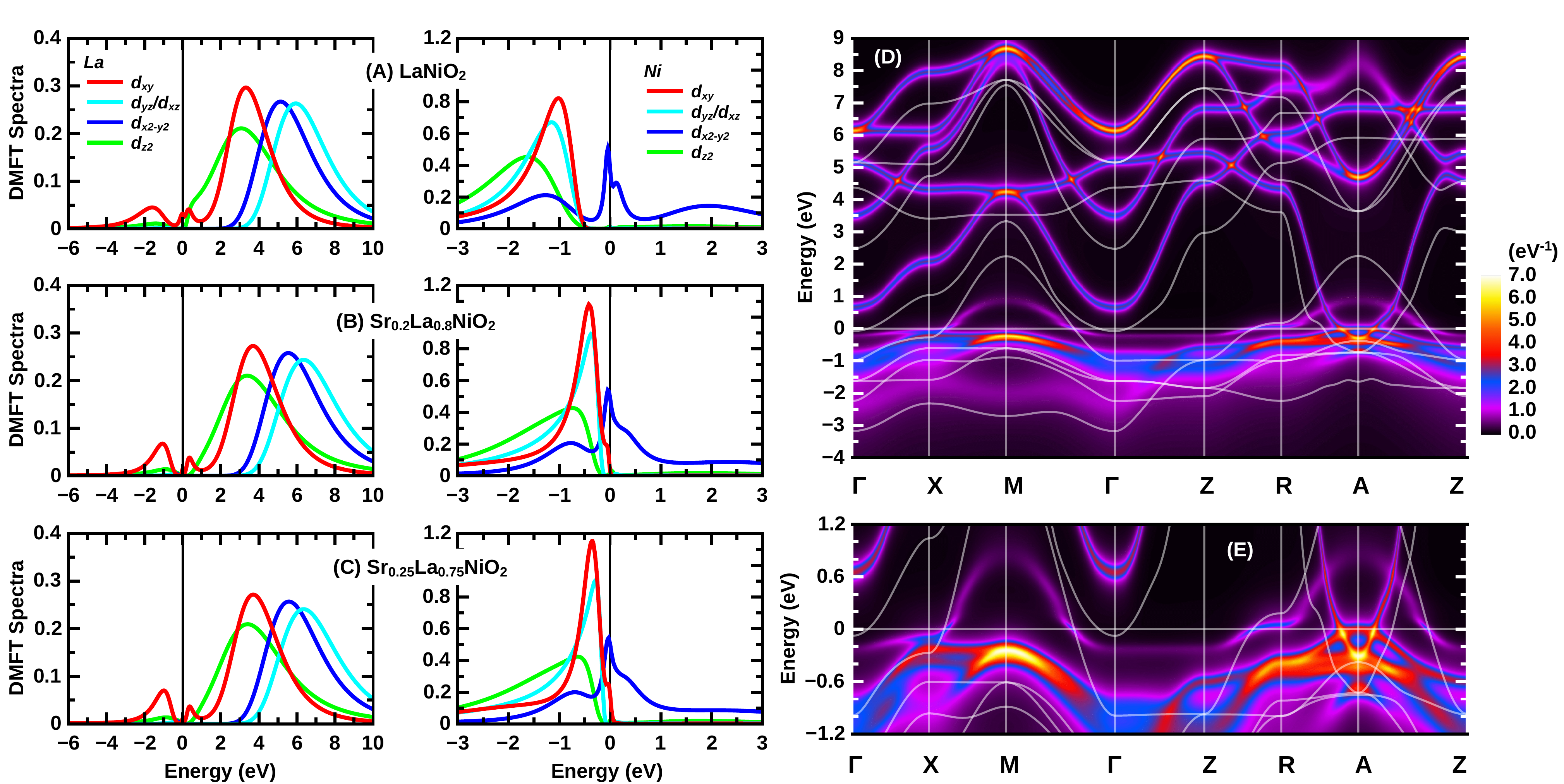}
\caption{
DMFT $k$-integrated (A-C) and $k$-resolved (D-E) spectral functions $A(\omega)$ and $A(k,\omega)$ of undoped LaNiO$_2$ (A), 20\%Sr-doped LaNiO$_2$ (Sr$_{0.2}$La$_{0.8}$NiO$_2$) (B), and 25\%Sr-doped LaNiO$_2$ (Sr$_{0.25}$La$_{0.75}$NiO$_2$) (C). 
The  $k$-resolved spectral function $A(k,\omega)$ of La$_{0.8}$Sr$_{0.2}$NiO$_2$ is shown in (D); (E) is a  zoom-in of (D). Data partially from~\cite{Si2019,Kitatani2020}.} 
\label{FigDFTDMFT}
\end{figure*}

Fig.~\ref{FigDFTDMFT} shows the  DFT+DMFT spectral function for Sr$_{x}$La$_{1-x}$NiO$_2$ from 0\% to 25\% Sr-doping. In these calculations~\cite{Kitatani2020} all Ni 3$d$ and all La 5$d$ orbitals have been taken into account in {a \textsc{wien2wannier}~\cite{kunevs2010wien2wannier} projection supplemented by interactions  calculated within the} constrained random phase approximation (cRPA)~\cite{Si2019} to be 
$U'= 3.10\,$eV ($2.00\,$eV) and Hund's exchange  $J=0.65\,$eV  (0.25\,eV) for Ni (La). On a technical note, the DMFT self-consistency equations~\cite{RevModPhys.68.13}  have been solved here at  room temperature (300\,K) by continuous-time quantum Monte Carlo simulations in the hybridization expansions~\cite{RevModPhys.83.349} using the
 \textsc{w2dynamics} implementation~\cite{PhysRevB.86.155158,w2dynamics2018} and the   maximum entropy code of \textsc{ana\_cont}~\cite{Kaufmann2021} for analytic continuation.

Clearly, 
Fig.~\ref{FigDFTDMFT} indicates that for up to 20\% Sr-doping the other Ni 3$d$ orbitals besides the  3$d_{x^2-y^2}$ orbital are not relevant for the low-energy physics of Sr$_{x}$La$_{1-x}$NiO$_2$: they are fully occupied below the Fermi energy. With doping, these other Ni 3$d$ orbitals however shift more and more upwards in energy.  At  {around}  25\% Sr-doping they {touch} the Fermi energy and hence a multi-orbital Ni description becomes necessary {at larger dopings}. That is, between 20\% and 30\% Sr-doping  the physics of Sr$_{x}$La$_{1-x}$NiO$_2$ turns from  single- to  multi-orbital. In the case of Sr$_{x}$Nd$_{1-x}$NiO$_2$, this turning point is at slightly larger doping~\cite{Kitatani2020}.
Later, in Section~\ref{Sec:topH}, we will see that for {the} Ni 3$d^8$ configuration, which in Section~\ref{Sec:topH} is induced by topotactic hydrogen and here would be obtained for 100\% Sr doping,  the two holes in the Ni 3$d$ orbitals form a spin-1
and occupy two orbitals:  3$d_{x^2-y^2}$ and 3$d_{3z^2-r^2}$. {I}n Fig.~\ref{FigDFTDMFT} we see at 30\% doping the first steps into this direction. Importantly, within the superconducting doping regime which noteworthy is {\em below} 24\% Sr-doping for Sr$_{x}$Nd$_{1-x}$NiO$_2$~\cite{Li2020,zeng2020} and
21\% for  Sr$_{x}$La$_{1-x}$NiO$_2$~\cite{Osada2021}, a single   3$d_{x^2-y^2}$ Ni-orbital is sufficient for the low-energy modelling.
A one-band Hubbard model description based on DMFT calculations was also concluded in~\cite{Karp2020} for the undoped parent compound.

DFT+DMFT calculations by Lechermann~\cite{Lechermann2019,Lechermann2020} stress, on the other hand, the relevance of the  3$d_{3z^2-r^2}$ orbital. Let us note that also in~\cite{Lechermann2019} the number of holes in the 3$d_{3z^2-r^2}$ orbital is considerably less than in the  3$d_{x^2-y^2}$. However, for low Sr-doping also a {small} quasiparticle peak develops for the  3$d_{3z^2-r^2}$ band~\cite{Lechermann2019,Lechermann2020}.  An important difference to ~\cite{Si2019,Kitatani2020} is that the 5$d$ Coulomb interaction has been taken into account in ~\cite{Si2019,Kitatani2020} {and that the Coulomb interaction of~\cite{Lechermann2019} is substantially larger}. The {5$d$ Coulomb interaction} pushes the $\Gamma$ pocket above the Fermi energy (see next paragraph).
As much of the holes in the  3$d_{3z^2-r^2}$ orbital stem from the admixture of this orbital to the $\Gamma$ pocket, this difference is very crucial for the occupation of the   3$d_{3z^2-r^2}$ orbital in some calculations~\cite{Lechermann2020,Adhikary2020}. On the other hand, in $GW$+extended DMFT calculations by Petocchi {\em et al.}~\cite{Petocchi2020}, the  3$d_{3z^2-r^2}$ orbital is pushed to the Fernmi energy for large $k_z$ instead, i.e.,  around the $R$, $Z$ and $A$ point. Except for this large $k_z$ deviation, the Fermi surface, the effective mass
of the Ni 3$d_{x^2-y^2}$ orbital etc.\ of  \cite{Petocchi2020} are similar to our calculation~\cite{Kitatani2020,Worm2021c}.

First experimental hints on the (ir)relevance of the   3$d_{3z^2-r^2}$ can  be obtained from resonant inelastic x-ray scattering (RIXS) experiments~\cite{Hepting2020,Lu2021}.  Higashi {\em et al.}~\cite{Higashi2021}
analyzed  these RIXS data by comparison with DFT+DMFT and obtained good agreement with experiment.
They conclude that NdNiO$_2$ is slightly doped away from 3$d^9$ because of a small self-doping from the Nd 5$d$ band, that  only the  3$d_{x^2-y^2}$ Ni orbital (not the  3$d_{3z^2-r^2}$ orbital) is
 partially filled, and 
that  {the Ni-O hybridization  plays a less important role than for the cuprates.}

\paragraph*{$\Gamma$ pocket.} A feature clearly present in  DFT calculations for the nickelate parent compounds LaNiO$_2$ and  NdNiO$_2$ is the $\Gamma$ pocket, see Fig.~\ref{Fig:DFTbands}{(D)}. However, when the Coulomb interaction on the La or Nd sites is included{,} it is shifted upwards in energy. Furthermore, Sr-doping depopulates the Ni 3$d_{x^2-y^2}$   orbital as well as the $A$ and $\Gamma$ pocket, and thus also helps pushing the $\Gamma$ pocket above the Fermi energy. Clearly in the DFT+DMFT {$k$}-resolved spectrum of Fig.~\ref{FigDFTDMFT}, the $\Gamma$ pocket is above the Fermi energy.
To some exten{t} the presence or absence of the $\Gamma$ pocket also depends on the rare{-}earth cation. For NdNiO$_2$ we obtain a $\Gamma$ pocket for the undoped compound~\cite{Kitatani2020} which only shifts above the Fermi energy with Sr-doping in the superconducting region, whereas for  LaNiO$_2$  it is already above the Fermi level without Sr-doping. 
We can hence conclude that while there might be a $\Gamma$ pocket without Sr-doping, DFT+DMFT results suggest that it is absent in the superconducting doping regime.

Briefly after the discovery of superconductivity in nickelates, it has also been  suggested that the Nd 5$d$ orbitals of the pockets couple to
the Ni 3$d_{x^2-y^2}$ spin, giving rise to a  Kondo effect~{\cite{Zhang2019,Gu2020}}.  However, Table~\ref{table1} shows that the hybridization between the relevant  Ni 3$d_{x^2-y^2}$  and the most important La or Nd 5$d_{xy}$ and 5$d_{3z^2-r^2}$
 vanishes by symmetry. {Also the full 5 Ni and 5 Nd band DMFT calculation in Fig.~\ref{FigDFTDMFT} does not show a hybridization (gap) between $A$ pocket and Ni bands.} This {suggests} that the
$\Gamma$ and $A$ pocket are  decoupled from the  3$d_{x^2-y^2}$  orbitals. There is no hybridization and hence no Kondo effect.






\begin{table}
\begin{tabular}{c|c|c|c|c|c}
\hline
\hline
\\[-1.7ex]~   
LaNiO$_2$  & La 5$d_{xy}$ & La 5$d_{yz}$ & La 5$d_{xz}$ & La 5$d_{x^2-y^2}$ & La 5$d_{z^2}$  \\ \hline
\\[-1.7ex]~   
Ni 3$d_{x^2-y^2}$ (10-bands model, GGA) & 0.000  &  0.084 & -0.084 & -0.017  & 0.0000  \\

Ni 3$d_{x^2-y^2}$ (17-bands model, GGA) & 0.000  &  0.085  & -0.085  & -0.037 & 0.0000  \\

\hline
\hline
\\[-1.7ex]~   
NdNiO$_2$ & Nd 5$d_{xy}$ & Nd 5$d_{yz}$ & Nd 5$d_{xz}$ & Nd 5$d_{x^2-y^2}$ & Nd 5$d_{z^2}$  \\
\hline
\\[-1.7ex]~   
Ni 3$d_{x^2-y^2}$ (10-bands model, GGA open core) & 0.0000  &   0.0701 & -0.0701 & -0.0388   & 0.0000  \\

Ni 3$d_{x^2-y^2}$ (10-bands model, GGA) & 0.0000  &   0.0775 & -0.0775  & -0.0066    & 0.0000  \\

Ni $d_{x^2-y^2}$ (17-bands model, GGA) & 0.0000  & 0.0811   & -0.0811  & -0.0239 & 0.0000  \\
\hline
\hline
\end{tabular}
\caption{Hybridization (hopping amplitude in eV) between the partially occupied  Ni 3$d_{x^2-y^2}$ and the La/Nd 5$d$ orbitals~\cite{Kitatani2020}. {Here, the results are obtained from Wannier projections onto 17-bands (La/Nd-4$f$+La/Nd-5$d$+Ni-3$d$) and 10-bands  (La/Nd-5$d$+Ni-3$d$).}
\label{table1}}
\end{table}

\paragraph*{Nd 4$f$ orbitals.}
 Finally, the importance of the Nd 4$f$ orbitals
 has been  suggested in the literature. Treating these 4$f$ orbitals in DFT is not trivial, because DFT puts them in the vicinity of the Fermi level. This neglects that electronic correlations split the Nd 4$f$ into upper and lower Hubbard bands, as they form a local spin. This effect is beyond DFT. One way to circumvent this difficulty is to put the   Nd 4$f$ orbitals in the core instead of having them as valence states close to the Fermi energy. This is denoted as 
 ``GGA open core'' instead of standard  ``GGA'' 
 in Table~\ref{table1}. The localized Nd 4$f$ spins might in principle be screened through a Kondo effect. However, the hybridization
 of the Nd 4$f$ with the Ni  3$d_{x^2-y^2}$ orbital at the Fermi energy is extremely small, see Table~\ref{table2} and \cite{jiang2019electronic}. Hence, the  Kondo temperature is zero for all practical purposes. In spin-polarized DFT+$U$ there is instead a local exchange interaction between the Nd 4$f$ and the predominately Nd 5$d$ $\Gamma$ pocket~\cite{Choi2020}. However, as pointed out in the previous paragraph, the $\Gamma$ pocket is shifted above the Fermi level in the superconducting Sr-doping regime. Hence in~\cite{Kitatani2020}, we ruled out that the Nd 4$f$ are relevant for superconductivity.
 This has been spectacularly confirmed experimentally by     the discovery of superconductivity in nickelates  without $f$ electrons: Ca$_x$La$_{1-x}$NiO$_2$~\cite{Zeng2021}  and Sr$_x$La$_{1-x}$NiO$_2$~\cite{Osada2021} have  a similar $T_c$.

\begin{table}
\centering
\begin{tabular}{c|c|c|c|c|c|c|c}
\hline
\hline
\\[-1.7ex]~   
LaNiO$_2$ (GGA) & $f_{xz^2}$ & $f_{yz^2}$ & $f_{z^3}$ & $f_{x(x^2-3y^2)}$ & $f_{y(3x^2-y^2)}$ & $f_{z(x^2-y^2)}$ & $f_{xyz}$ \\
\hline
\\[-1.7ex]~   
Ni-$d_{x^2-y^2}$ & -0.0300   & 0.0300 & 0.0000 & -0.0851 & -0.0851 & -0.0203 & -0.0000 \\
\hline
\hline
\\[-1.7ex]~   
NdNiO$_2$ (GGA) & $f_{xz^2}$ & $f_{yz^2}$ & $f_{z^3}$ & $f_{x(x^2-3y^2)}$ & $f_{y(3x^2-y^2)}$ & $f_{z(x^2-y^2)}$ & $f_{xyz}$ \\
\hline
\\[-1.7ex]~   
Ni-$d_{x^2-y^2}$ & -0.0215  & 0.0215 & 0.0000 & -0.0612 & -0.0612 & 0.0160 & -0.0000   \\
\hline
\hline
\end{tabular}
\caption{Hybridization (hopping amplitude in eV) between the Ni 3$d_{x^2-y^2}$ and the Nd(La) 4$f$ orbitals, as  obtained from Wannier projections onto 17-bands (La/Nd-4$f$+La/Nd-5$d$+Ni-3$d$) including the  4$f$ as valence states in DFT(GGA)~\cite{Kitatani2020}.
\label{table2}}
\end{table}

\section{One-band Hubbard model plus reservoir}
        \label{Sec:fill}

Altogether this leaves us with  Fig.~\ref{Fig:Model}~B,D as the arguably simplest model for nickelate superconductors, consisting of  a strongly correlated
{Ni 3$d_{x^2-y^2}$ band} and an {$A$ pocket}. {This $A$ pocket is derived from the 
Nd 5$d_{xy}$ band which however crosses the Ni 3$d$ orbitals and hybridizes strongly with the 
Ni $t_{2g}$ orbitals so that at the bottom of the $A$ pocket, i.e., at the momentum $A$, it is made up primarily from Ni $t_{2g}$ whereas the Nd 5$d_{xy}$ contribution is here at a lower energy.
This makes the $A$ pocket  much more resistive to shifting up in energy than the $\Gamma$ pocket.}

{On the other hand the $A$ pocket  does not interact with the Ni 3$d_{x^2-y^2}$ band; i.e.,  does not hybridize in Table~\ref{table1}.} Hence, we can consider the {$A$ pocket} as a mere 
hole  {reservoir} which accommodates part of the holes induced by Sr doping, whereas the other part  goes into the correlated Ni 3$d_{x^2-y^2}$ band which is responsible for  superconductivity.
Fig.~\ref{Fig:doping} shows the thus obtained Ni 3$d_{x^2-y^2}$
occupation as a function of Sr-doping in the DFT+DMFT calculation with 5 Ni and 5 Nd(La) orbitals.

Note that NdNiO$_2$ shows for Sr-doping below about 10\% 
more holes in the  Ni 3$d_{x^2-y^2}$ orbital and a weaker dependence on the Sr-doping, since here the $\Gamma$ pocket is still active, taking away electrons from Ni but also {first} absorbing some of the holes from the Sr-doping until it is completely depopulated (shifted above the Fermi energy) before superconductivity sets in.

{In the subsequent one-band calculation, presented in the next paragraph, we employ the occupation from the Ni 3$d_{x^2-y^2}$ orbital as calculated in this full DMFT calculation with 5 Ni and 5 Nd orbitals. This accounts for the electron pocket in the DMFT calculation but also for minor hybridization effects between
  the Ni 3$d_{x^2-y^2}$ and 3$d_{3z^2-r^2}$ orbital, e.g., along the $\Gamma$-$X$ direction.
  In principle, this hybridization effect, which intermixes the orbital contribution to the bands, should not be taken into account in
  the one-band Hubbard model. This aims at modelling the {\em effective} 3$d_{x^2-y^2}$ band which  is crossing the Fermi level and which is predominantly Ni  3$d_{x^2-y^2}$ (but also has admixtures from the other orbitals because of the hybridization). In the case of
the  Ni  3$d_{x^2-y^2}$ orbital this hybridization is very weak \cite{Kitatani2020} (Supplemental Material) and can be neglected as a first approximation~\cite{Worm2021c} (Supplemental Material). For other  Ni 3$d$ orbitals this hybridization has a sizable effect on their respective occupation. For example
the Ni 3$d_{3z^2-r^2}$ orbital which is strongly hybridizing with the Nd  5$d_{3z^2-r^2}$ orbital only has an occupation of 1.85 electrons per site in our multi-orbital calculation~\cite{Worm2021c} (Supplemental material), whereas the {\em effective}  Ni  3$d_{3z^2-r^2}$ orbital including contributions from the hybridization is fully occupied with 2 electrons per site, as it is completely below the Fermi level.

\begin{figure*}[tb]\centering
\includegraphics[width=15cm]{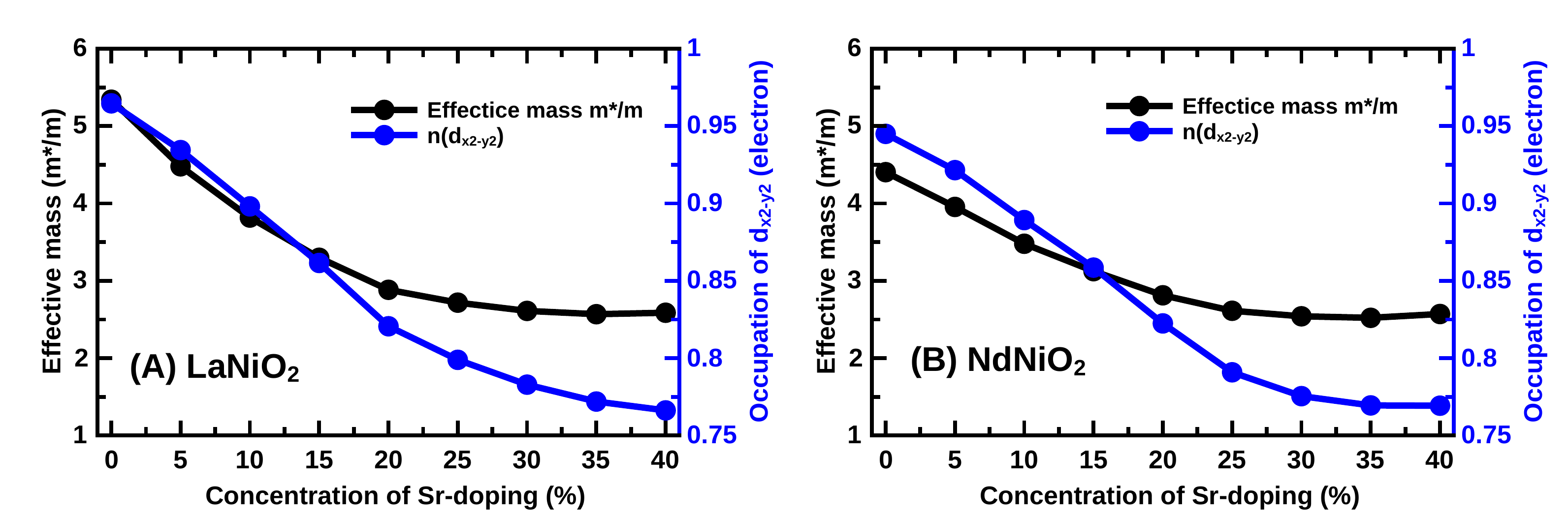}
\caption{Occupation of the Ni 3$d_{x^2-y^2}$ orbital [blue; right $y$-axis] and its  effective mass enhancement {$m^*/m=1/Z$}[black ; left $y$-axis in panels] vs.~Sr-doping for LaNiO$_2$ (A) and NdNiO$_2$ (B) as calculated in DFT+DMFT. From~\cite{Kitatani2020} (Supplemental material).}
\label{Fig:doping}
\end{figure*}

The hopping parameters for {the} Ni-3$d_{x^2-y^2}$ model from  a
one-band Wannier projection
are shown in  Table~\ref{table3}, and compared to that of the same orbital in a 10-band and 17-band Wannier projection. Here $t_{R_x,R_y,R_z}$ denotes the hopping by $R_i$ unit cells in the $i$ direction. That is, $t_{000}$ is the on-site potential, $t=-t_{100}$, $t'=-t_{110}$ and $t''=-t_{200}$ are the  nearest, next-nearest and next-next-nearest neighbor hopping;
 $t_z=-t_{001}$ is the hopping in the $z$-direction perpendicular to the NiO$_2$ planes. The hopping parameters are strikingly similar for LaNiO$_2$ and  NdNiO$_2$ and the different Wannier projections.

Besides the doping from Fig.~\ref{Fig:doping} {and} the hopping for the one-band Wannier projection from  Table~\ref{table3}, we only need the interaction parameter for doing realistic one-band Hubbard model calculations for nickelates.
In cRPA for a single 3$d_{x^2-y^2}$ orbital  one obtains $U=2.6\,$eV \cite{Nomura2019,Hirofumi2019} 
at zero frequency. But the cRPA interaction  has a strong frequency dependence because of the screening of all the other Ni and Nd(La) orbitals close by. To mimic this frequency
dependence, the static $U$ parameter needs to be slightly increased. Expertise from many DFT+DMFT calculations for transition metal oxides shows that it typically needs to be about 0.5$\,$eV larger, so that  $U=3.2\,$eV$=8t$ is reasonable. Altogether, this defines a one-band Hubbard model for nickelates at various dopings. For the conductivity and other {transport} properties, the $A$ pocket may be relevant as well, but superconductivity should arise from the correlated
3$d_{x^2-y^2}$ band that is hardly coupled to the $A$ pocket.

\begin{table}[tb]
\centering
\begin{tabular}{c|c|c|c|c|c|c}
\hline
\hline
\\[-1.7ex]~   
LaNiO$_2$ (GGA) & $t_{000}$&$t_{100}$&$t_{001}$&$t_{110}$&$t_{200}$&$t_{210}$  \\
\hline
\\[-1.7ex]~   
1-band (Ni-$d_{x^2-y^2}$) &  0.2689 & -0.3894 & -0.0362 & 0.0977 & -0.0465 & -0.0037 \\

10-bands (La-$d$+Ni-$d$)   &  0.2955 & -0.3975 & -0.0458 & 0.0985 & -0.0491 & 0.0000 \\

17-bands (La-$f$+La-$d$+Ni-$d$)   &   0.3514  &  -0.3943 & -0.0239 & 0.0792  & -0.0422  & -0.0008  \\
\hline
\hline
\\[-1.7ex]~   
NdNiO$_2$ (GGA open core) & $t_{000}$&$t_{100}$&$t_{001}$&$t_{110}$&$t_{200}$&$t_{210}$  \\
\hline
\\[-1.7ex]~   
1-band (Ni-$d_{x^2-y^2}$) &   0.3058  & -0.3945  &  -0.0336 & 0.0953  & -0.0471 &  -0.0031 \\

10-bands (Nd-$d$+Ni-$d$)   &  0.3168   &  -0.3976 & -0.0389  &  0.0949  &  -0.0480  & -0.0008   \\
\hline
\hline
\end{tabular}
\caption{Major hopping elements (in units of eV)  of the Ni-3$d_{x^2-y^2}$ orbital from 1-band (Ni-3$d_{x^2-y^2}$), 10-bands (La/Nd-$d$+Ni-$d$) and 17-bands (La/Nd-$f$+La/Nd-$d$+Ni-$d$) Wannier projections.  The 
DFT-relaxed lattice parameters are: LaNiO$_2$  ($a=b=3.88\,$\AA{}, $c=3.35\,$\AA{}),  NdNiO$_2$  ($a=b=3.86\,$\AA{}, $c=3.24\,$\AA{})~\cite{Kitatani2020}.
\label{table3}}
\end{table}

\section{Non-local correlations and superconducting phase diagram}
        \label{Sec:DGA}
DFT provides  a first picture of the relevant orbitals, and DMFT adds to this effects of strong local correlations such as the splitting into Hubbard bands, the formation of a quasiparticle peak and  correlation-induced orbital shifts such as the upshift of the $\Gamma$ pocket. However, at low temperatures non-local correlations give rise to additional effects. Relevant are here: the emergence of strong spin fluctuations {and} their impact on the spectral function and superconductivity.

For including such non-local correlations,  diagrammatic extensions of DMFT such as the dynamical vertex approximation (D$\Gamma$A)~\cite{Toschi2007,Katanin2009,Galler2016,RMPVertex} have been proven extremely powerful. Such calculations are possible down to the temperatures of the superconducting phase transition, in the correlated regime and for very large lattices so that the long-range correlations close to a phase transition can be properly described. Even (quantum) critical exponents can be calculated~\cite{Rohringer2011,Antipov2014,Schaefer2016,Schaefer2019}. D$\Gamma$A
 has proven reliable compared with  numerically exact calculations where these are possible~\cite{Schaefer2021}, and in particular provide for a more accurate determination of $T_c$~\cite{Kitatani2019} since the full local frequency dependence of the {two-particle} vertex is included. {Such local frequency dependence can affect even the non-local pairing through spin fluctuations.
}
In, e.g., RPA this frequency dependence and the suppression of the {pairing} vertex for small frequencies can  only be improperly mimicked by (quite arbitrarily) adjusting the static $U$.

This simple one-band Hubbard model in D$\Gamma$A has been the basis for calculating the phase diagram $T_c$ vs.~Sr-doping  in Fig.~\ref{FigTc}~\cite{Kitatani2020}. At the time of the calculation only a single experimental $T_c$ at 20\% Sr-doping was available~\cite{li2019superconductivity}.
The physical origin of the superconductivity in these calculations are strong spin fluctuations which form the pairing glue for high-temperature superconductivity.  Charge fluctuations are much weaker; the electron-phonon {coupling} has not been  considered and is also too weak for transition metal oxides to yield high-temperature superconductivity.
The theoretical $T_c$   in Fig.~\ref{FigTc} at 20\% {doping} was from the very beginning slightly larger than in experiment. Most likely this is because in the ladder  D$\Gamma$A~\cite{Katanin2009,RMPVertex} calculation of $T_c$ the spin fluctuations are first calculated and then enter the superconducting particle-particle channel~\cite{Kitatani2019}. {This} neglects the feedback effect of these particle-particle fluctuations on the self-energy and the spin fluctuations, which may in turn suppress the tendency towards superconductivity somewhat. Such effects would be only included in a more complete parquet D$\Gamma$A calculation~\cite{Valli2015,Li2016,Li2017}. {Also, the ignored weak three-dimensional dispersion will suppress Tc.} Let us note that 
antiferromagnetic spin fluctuations have recently been observed experimentally \cite{Lu2021,Cui2021}.

Given the slight  overestimation of $T_c$ from the very beginning, the agreement with the subsequently obtained experimental $T_c$ vs.~Sr-doping $x$ phase diagram~\cite{Li2020,zeng2020}  in Fig.~\ref{FigTc} is astonishingly good. We further see that the superconducting doping regime also concurs with the doping regime where a one-band Hubbard model description is possible  for Sr$_x$Nd$_{1-x}$NiO$_2$, as concluded from a full DFT+DMFT calculation for 5 Ni plus 5 Nd bands. This regime is marked
dark blue in  Fig.~\ref{FigTc} and, as already noted, extends to somewhat larger dopings~\cite{Kitatani2020} than for  Sr$_x$La$_{1-x}$NiO$_2$ shown in
Fig.~\ref{FigDFTDMFT}. Concomitant with this is the fact that the experimental superconducting doping range for  Sr$_x$La$_{1-x}$NiO$_2$ extends  to a larger $x$  than for   Sr$_x$Nd$_{1-x}$NiO$_2$. For dopings  larger than the dark blue regime in  Fig.~\ref{FigTc},  two Ni 3$d$ bands  need to be included. As we will show in the next Section, this completely changes the physics and is not favorable for superconductivity. For dopings smaller than the dark blue regime in  Fig.~\ref{FigTc}, on the other hand, the $\Gamma$ pocket  may become relevant for Sr$_x$Nd$_{1-x}$NiO$_2$, {as well as} its exchange coupling to the 4$f$  moments. 

 Our theoretical calculations also reveal ways to enhance $T_c$. Particularly promising is to enhance the hopping parameter $t$.  This enhances $T_c$ because (i) $t$ sets the energy scale of the problem and a larger $t$ means a larger $T_c$ if $U/t$, $t'/t$, $t''/t$ and doping are kept fixed. Further the ratio $U/t=8$ for nickelates is not yet optimal. Indeed,  (ii) {a} somewhat smaller ratio $U/t$ would imply a larger $T_c$ at fixed $t$~\cite{Kitatani2020}. Since the interaction $U$ is local it typically varies much more slowly when, e.g., applying compressive strain or pressure and can be assumed to be constant as a first approximation {(for secondary effects,} {see \cite{Ivashko2019,Upressure})}. Thus compressive strain or pressure enhance (i) $t$ and reduce (ii) $U/t$. Both effects enhance $T_c$. This prediction made in~\cite{Kitatani2020}
has been confirmed experimentally: applying pressure of 12\,GPa increases $T_c$ from 18$\,$K to 31$\,$K in Sr$_{0.18}$Pr$_{0.82}$NiO$_2$ \cite{Wang2021}. This is so far the record $T_c$ for nickelates, and there are  yet no signs for a saturation or maximum, indicating even higher $T_c$'s are possible at higher pressures.

Alternatives to enhance $t$ are  {(1)} to substitute the  SrTiO$_3$ substrate by a substrate with smaller in-plane lattice constants since the nickelate film in-plane axis parameters will be locked to that of the substrate. Further, one can {(2)} replace 3$d$ Ni by 4$d$ Pd, i.e.\ try to synthesize Nd(La)PdO$_2$~{\cite{Motoaki2019}}. Since the  Pd  4$d$ orbitals are more extended than the 3$d$ Ni orbitals this should enhance $t$ as well.

\begin{figure}[t]
  \begin{minipage}{.5\textwidth}
    \vspace{-.5cm}
    
    \includegraphics[width=.87\textwidth]{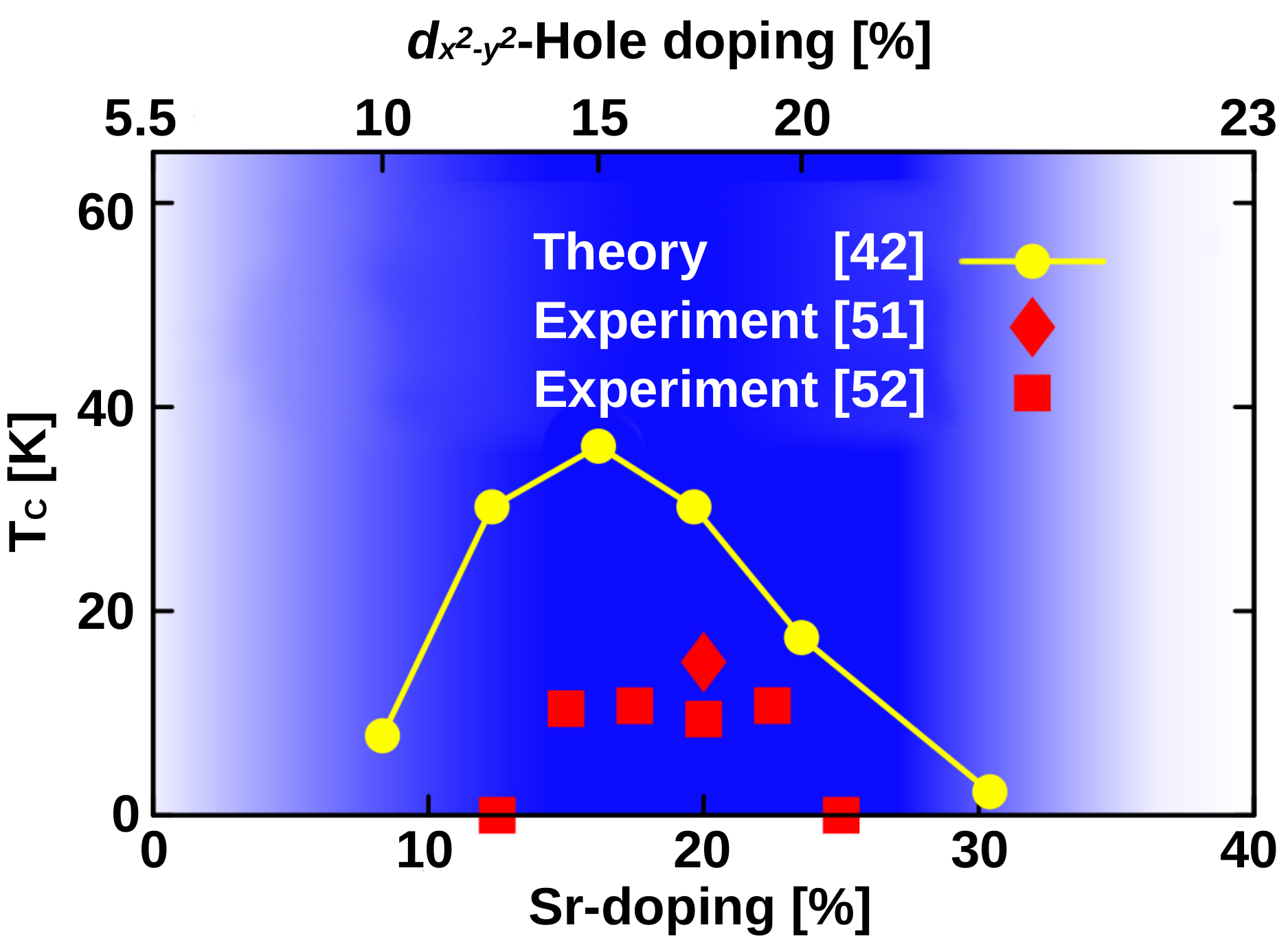}
    \vspace{-.2cm}
\end{minipage}\hfill
  \begin{minipage}{.49\textwidth}
    \caption{Superconducting phase diagram $T_c$ vs.~$x$  for Sr$_x$Nd$_{1-x}$NiO$_2$ as predicted by D$\Gamma$A~\cite{Kitatani2020} and experimentally confirmed  {\it a posteriori} in~\cite{Li2020} and~\cite{zeng2020}. {\it A priori}, i.e., at the time of the calculation only one experimental data point~\cite{li2019superconductivity} was available. \label{Fig:PD}  \label{FigTc} For such a difficult to determine quantity as the superconducting $T_c$ and without adjusting  parameters, the accuracy  is astonishing. The bottom $x$-axis shows the Sr-doping and the top $x$-axis the calculated  hole doping of the 3$d_{x ^2-y^2}$ band according to Fig.~\ref{Fig:doping}.
      Adjusted from \cite{Kitatani2020}.}
\end{minipage}
\end{figure}

Next, we turn to the D$\Gamma$A spectra, more precisely Fermi surfaces,  in Fig.~\ref{Fig:DGA}.  Here, beyond quasiparticle renormalizations of DMFT, non-local spin fluctuations can further impact the spectrum. Shown is only the spectral function of the Hubbard model,
describing the 3$d_{x^2-y^2}$ band. Please {keep in mind}, that on top of the Fermi surface in  Fig.~\ref{Fig:DGA}, there is also a weakly correlated $A$ pocket. As one can see  in Fig.~\ref{Fig:DGA} antiferromagnetic spin-fluctuations lead to a pseudogap at the antinodal momenta $(\pm \pi,0)$ $(0, \pm \pi)$ if the filling  of  the 3$d_{x^2-y^2}$ band is close to half-filling. Indeed
$n_{3d_{x^2-y^2}}=0.95$ is the filling for the undoped parent compound  NdNiO$_2$ where  the $A$- and $\Gamma$ pocket {have} taken 5\% of the electrons away from  the Ni 3$d_{x^2-y^2}$ band. A Sr-doping of 20\% is {in-between} $n_{3d_{x^2-y^2}}=0.85$ and $n_{3d_{x^2-y^2}}=0.8$, {see Fig.~\ref{Fig:doping}}. Comparing these theoretical predictions with the experimental Fermi surface, even the ${\mathbf k}$-integrated spectrum{,} is very much {sought} after. However, here we face the difficulty that the superconducting samples require a SrTiO$_3$ capping layer or otherwise may oxidize out of vacuum. This hinders photoemission spectroscopy (PES) experiments as these are extremely surface sensitive. Hitherto PES is only available without capping layer for Sr$_x$Pr$_{1-x}$NiO$_2$~\cite{Chen2021}. These show a surprisingly  low spectral density at the Fermi energy
despite the metallic behavior of the doped system, raising  the question of how similar these films are to the superconducting films.

\begin{figure*}[t]
  \begin{minipage}{.69\textwidth}
\includegraphics[width=11.8cm]{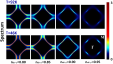}
\end{minipage}\hfill
  \begin{minipage}{.3\textwidth}
\caption{D$\Gamma$A {$k$}-resolved spectrum at the Fermi energy for $T=0.02t=92\,$K (upper panels) and  $T=0.01t=46\,$K (lower panels) and four different dopings $n_{d_{x^2-y^2}}$ of the Ni-3$d_{x^2-y^2}$ band (left to right). From \cite{Kitatani2020}. \label{Fig:DGA}}
  \end{minipage}
\end{figure*}

\section{Topotactic hydrogen: turning the electronic structure upside down}
\label{Sec:topH}
The fact that it took 20 years from the theoretical prediction of superconductivity in rare earth nickelates to the experimental realization already suggests that the synthesis is far from trivial. This is because nickel has to be in the unusually low oxidation state Ni$^{+1}$. The recipe of success for nickelate superconductors is a two step process
\cite{Lee2020}: First doped perovskite films
Sr(Ca)$_x$Nd(La,Pr)$_{1-x}$NiO$_3$ films are deposited on a SrTiO$_3$ substrate by pulsed laser deposition. Already this first step is far from trivial, {not least} because the doped material has to be deposited with homogeneous Sr(Ca) concentration. 
Second, Sr(Ca)$_x$Nd(La,Pr)$_{1-x}$NiO$_3$ needs to be reduced to
Sr(Ca)$_x$Nd(La,Pr)$_{1-x}$NiO$_2$. To this end,
the  reducing agent CaH$_2$ is employed. Here, the problem is that
this reduction might be incomplete with excess oxygen remaining or that hydrogen from CaH$_2$  is topotactically intercalated in the Sr(Ca)$_x$Nd(La,Pr)$_{1-x}$NiO$_2$ structure. A particular difficulty is that the light hydrogen is experimentally hard to detect, e.g., evades conventional x-ray structural detection.

In \cite{Si2019}, we studied the possibility to intercalate hydrogen, i.e., to synthesize unintendedly Sr$_x$Nd(La)$_{1-x}$NiO$_2$H instead of  Sr$_x$Nd(La)$_{1-x}$NiO$_2$.  For the reduction of, e.g., SrVO$_3$ with CaH$_2$  it is well established that SrVO$_2$H may be obtained as the end product~\cite{katayama2016epitaxial}. Both possible end products are visualized in
Fig.~\ref{Fig:topH}.  The extra H, takes away one more electron from the Ni sites. Hence, we have two holes on the Ni sites which in a local picture are distributed to two orbitals and form a spin-1, due to 
Hund's exchange.

\begin{figure}[t]
\begin{center}
\includegraphics[width=13.8cm]{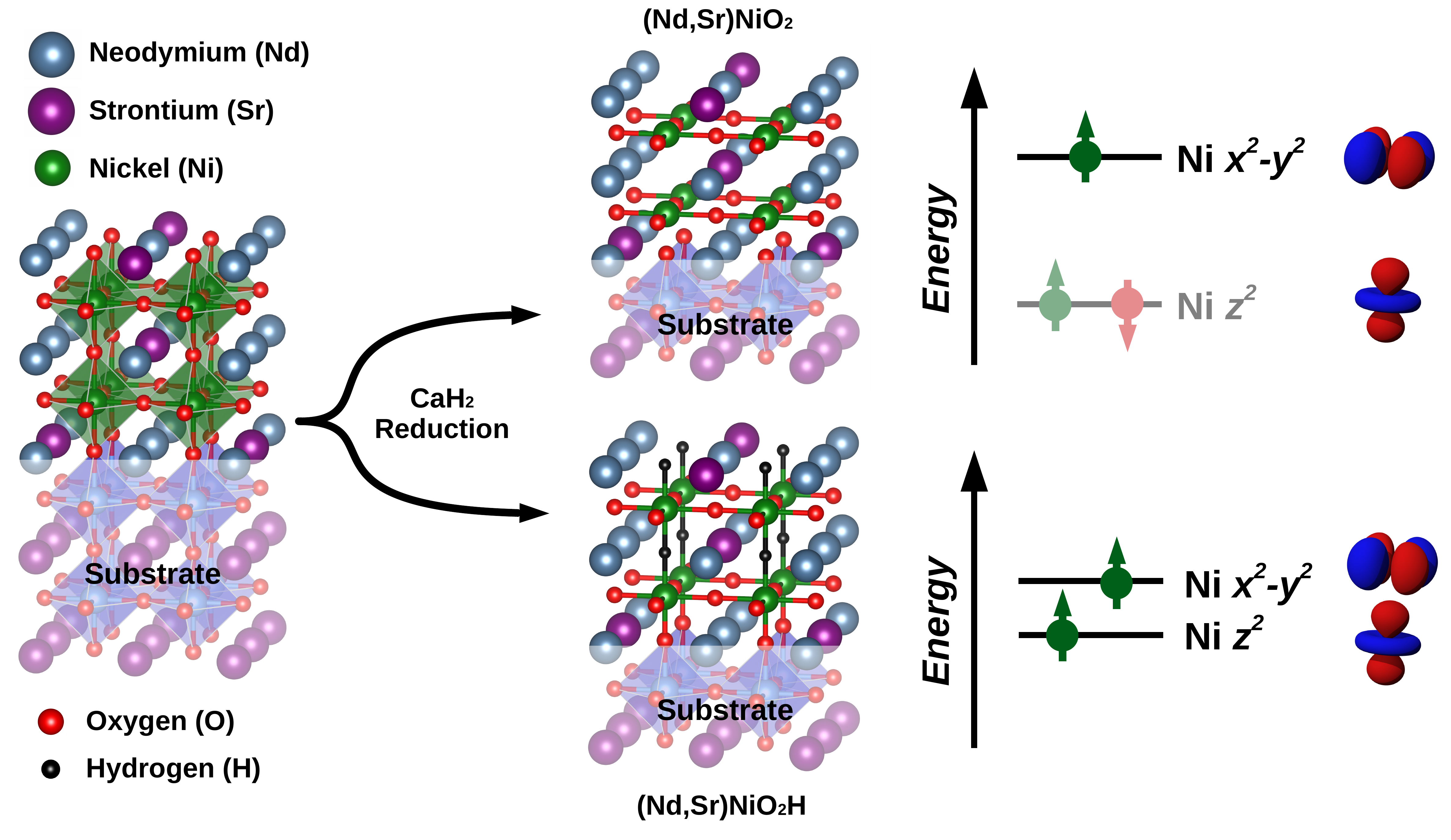}
\end{center}
\caption{The reduction of Sr$_x$Nd(La)$_{1-x}$NiO$_3$ with CaH$_2$ may result not only in the pursued end product Sr$_x$Nd(La)$_{1-x}$NiO$_2$ but also in Sr$_x$Nd(La)$_{1-x}$NiO$_2${H}, where H atoms occupy the vacant O sites between the layers. This has dramatic consequences for the electronic structure. In a first, purely local picture, visualized on the right side, we have
instead of Ni 3$d^9$ with one-hole in the    3$d_{x^2-y^2}$ orbital two holes in the    3$d_{x^2-y^2}$ and 3$d_{3z^2-r^2}$ orbital forming a local spin-1. \label{Fig:topH} }
\end{figure}

The first question is how susceptible the material is to bind topotactic H. To answer this question, one can calculate the binding energy E($AB$NiO$_2$) + 1/2 E(H$_2$)-
E($AB$NiO$_2$H)  in DFT~\cite{Si2019,Malyi2021}. The result is shown in Fig.~\ref{Fig:topHE}, which clearly shows that early transition metal oxides are prone to intercalate hydrogen, whereas for cuprates the infinite layer compound without H is more stable. Nickelates are in-between. For the undoped compounds NdNiO$_2$, and even a bit more for LaNiO$_2$, it is favorable to intercalate H. However for the Sr-doped nickelates the energy balance is inverted. Here, it is unfavorable to bind hydrogen.

Let us emphasize that this is only the enthalpy  balance. In the actual synthesis also the reaction kinetics matter, and the entropy which is large for the $H_2$ gas. Nonetheless, this shows that undoped nickelates are very susceptible to topotactic H. {This possibly means that, experimentally, not}
a complete H-coverage as in $AB$NiO$_2$H of  Fig.~\ref{Fig:topHE} {is realized}, but some  hydrogen may remain in the nickelates because of an incomplete reduction with CaH$_2$. 
Indeed hydrogen remainders have later been detected  experimentally by nuclear magnetic resonance (NMR) spectroscopy, and they have even been employed to analyze the antiferromagnetic spin fluctuations~\cite{Cui2021}.

\begin{figure}[t]
  \begin{minipage}{.59\textwidth}
\includegraphics[width=9.cm]{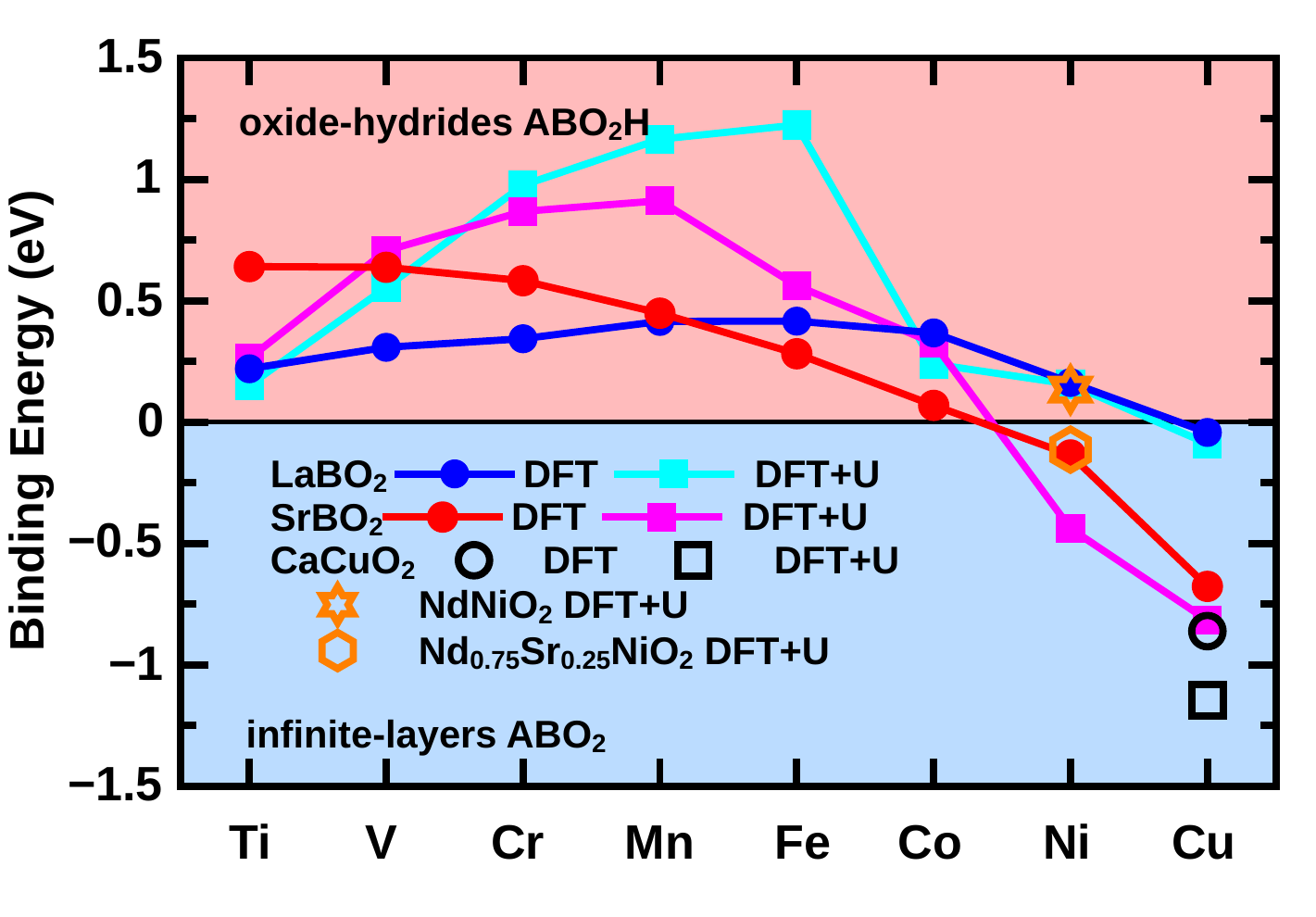}
\end{minipage}\hfill
  \begin{minipage}{.4\textwidth}
\caption{ Binding energy for topotactical H as calculated by non-spin-polarized DFT and spin-polarized DFT+U for various transition metals $B$ (x-axis).  Positive binding energies indicate  $AB$NiO$_2$H is energetically favored; for negative binding energies  $AB$NiO$_2$ is more stable. The data for non-spin-polarized DFT calculations are from~\cite{Si2019}. \label{Fig:topHE}}
  \end{minipage}
\end{figure}

Now that we have established that remainders of hydrogen can be expected for nickelates at low doping, the question is how this affects the electronic structure. The local picture of Fig.~\ref{Fig:topH} already suggested a very different electronic configuration. This is further corroborated by DFT+DMFT calculations for LaNiO$_2$H presented in  Fig.~\ref{Fig:topHA}. {Here,} the DFT {band structure} shows a metallic behavior with two orbitals, Ni   3$d_{x^2-y^2}$ and 3$d_{3z^2-r^2}$, crossing the Fermi level. There are no rare{-}earth electron pockets any longer. {Thus} we have an undoped Ni 3$d^8$ configuration without Sr-doping. If electronic correlations are included in DMFT, the DFT bands split into two sets of Hubbard bands. Above the Fermi level one can identify  the upper
 3$d_{x^2-y^2}$ and 3$d_{3z^2-r^2}$ Hubbard band bewlow the flat $f$ bands in  Fig.~\ref{Fig:topHA},  with quite some broadening because of the electronic correlations.
 The lower Hubbard bands intertwine with the Ni $t_{2g}$ orbitals below the Fermi energy.

 Even if we dope LaNiO$_2$H this electronic structure is not particularly promising for superconductivity. First, it is not two-dimensional  because of the 3$d_{3z^2-r^2}$ orbitals, which make the system more three-dimensional. More specifically, there is a considerable
 hopping process from Ni 3$d_{3z^2-r^2}$ via H to the Ni 3$d_{3z^2-r^2}$ on the {vertically} adjacent layer, as evidenced in  Fig.~\ref{Fig:topHA} by the DFT dispersion of this band in the $\Gamma$-Z direction, the other  3$d_{x^2-y^2}$ band is (as expected) flat in this direction. Second, the tendency to form local magnetic  moments of spin-1 counteracts the formation of  Cooper pairs from  two spin-1/2's. Hence, altogether,   we expect topotactic H to prevent high-temperature superconductivity.

\begin{figure}[t]
  \begin{minipage}{.69\textwidth}
\includegraphics[width=12cm]{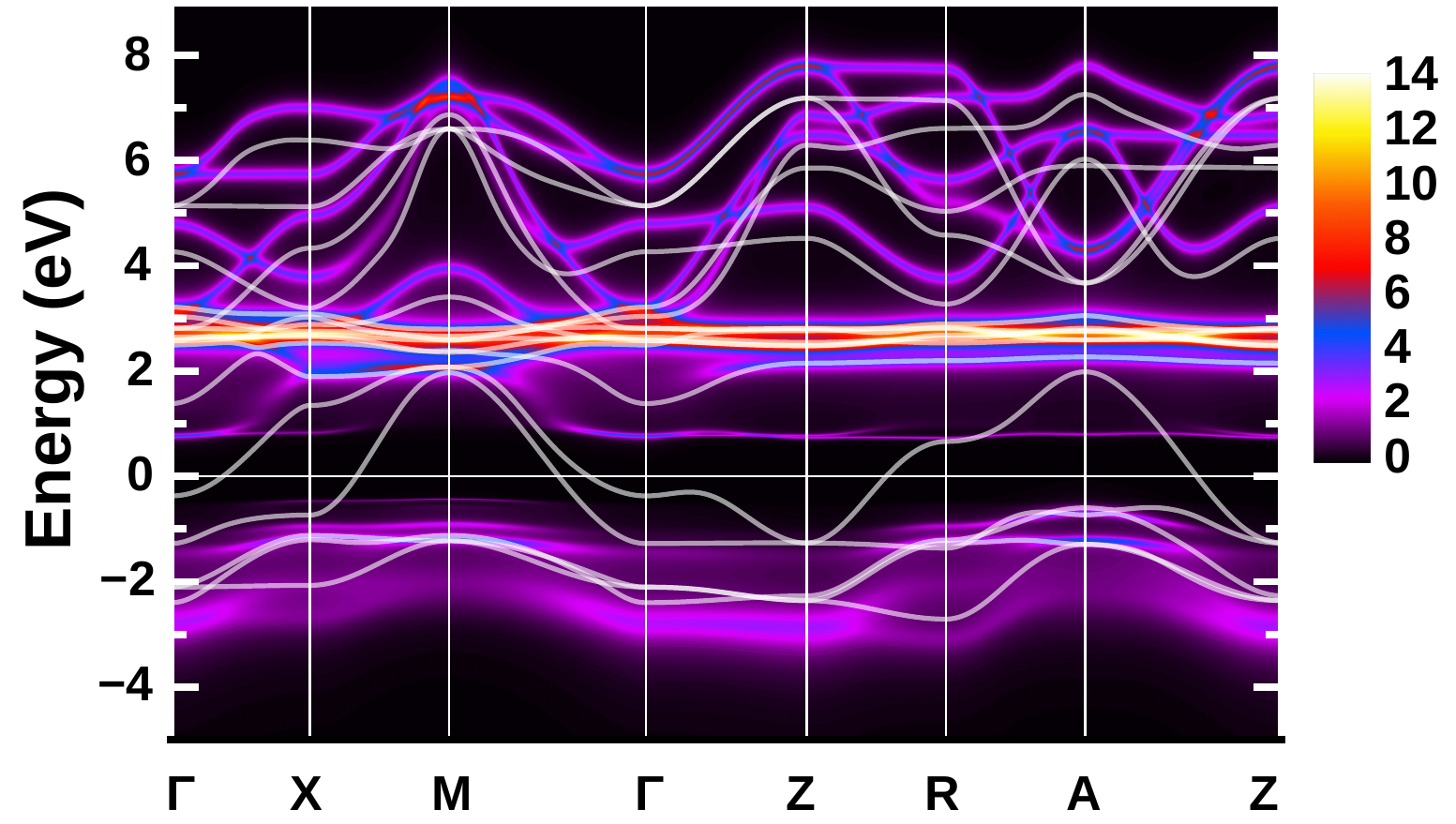}
\end{minipage}\hfill
  \begin{minipage}{.3\textwidth}
    \caption{ DFT (white lines) and DMFT (color bar) $k$-resolved spectral function for LaNiO$_2$H. Here, a model with La-d+La-f+Ni-d was employed, which is beyond the Ni-d only and La-d+Ni-d models shown in~\cite{Si2019}.\label{Fig:topHA} }
  \end{minipage}
\end{figure}

\section{Conclusion}
        \label{Sec:conclusio}
        In this paper, we have discussed the physics of nickelate superconductors from the perspective of a  one-band Hubbard model for the Ni  3$d_{x^2-y^2}$ band plus an $A$ pocket. Because of symmetry, this $A$ pocket does not hybridize with the  3$d_{x^2-y^2}$ band and merely acts  as a decoupled electron reservoir. Hence, once the filling of
        the  3$d_{x^2-y^2}$ band is calculated as a function of Sr or Ca doping in Sr(Ca)$_x$Nd(La,Pr)$_{1-x}$NiO$_2$, we can, for many aspects, concentrate on the physics of the thus doped Hubbard model. This includes antiferromagnetic spin fluctuations and the onset of superconductivity. Other physical properties such as transport and the Hall conductivity  depend as a matter of course also on the $A$ pocket. This is in stark contrast to the cuprates, where the oxygen $p$ orbitals are much closer to the Fermi level so that we have a charge{-}transfer insulator that needs to be modeled by the more complex Emery model.

        The  one-band Hubbard model picture for nickelates was put forward early on for nickelates~\cite{Wu2019,Motoaki2019,Karp2020,Kitatani2020} and its proper doping including correlation effects has been calculated in~\cite{Kitatani2020}.  This picture has been confirmed by many experimental observations so far. The Nd 4$f$ states are from the theoretical perspective irrelevant  because they form a local spin and barely  hybridize with the  3$d_{x^2-y^2}$ band. This has been   confirmed experimentally by the observation of superconductivity in Sr(Ca)$_x$La$_{1-x}$NiO$_2$. The minor importance of the other Ni 3$d$ orbitals, in particular the  3$d_{3z^2-r^2}$ orbital, is indicated through the careful analysis~\cite{Higashi2021} of RIXS data~\cite{Hepting2020,Lu2021}.  Not confirmed experimentally is hitherto the prediction that the $\Gamma$ pocket is shifted above the Fermi level in the superconducting doping regime.

        Strong evidence for the one-band Hubbard model picture is the prediction of the superconducting    phase diagram~\cite{Kitatani2020}, confirmed experimentally in~\cite{Li2020} and~\cite{zeng2020}.  A further prediction was the increase of $T_c$ with pressure or compressive strain~\cite{Kitatani2020} which was subsequently found in experiment with a record $T_c=31\,$K for nickelates under pressure~\cite{Wang2021}. The strength of antiferromagnetic spin-fluctuations as obtained in RIXS~\cite{Lu2021} also roughly agrees with that of the calculation~\cite{Kitatani2020}.
        Altogether, this gives us quite some confidence in the one-band Hubbard model scenario, {which even allowed for a} rough calculation of $T_c$. Notwithstanding, further theoretical calculations, in particular including non-local correlations also in a realistic multi-orbital {setting~\cite{RMPVertex,Galler2016,Tomczak2017}}, are eligible. On the experimental side more detailed, e.g., ${\mathbf k}$-resolved information is desirable as are further close comparisons between experiment and theory.

          A good analysis of the quality of the samples is also mandatory, especially against the background that superconducting nickelates have been extremely difficult to synthesize. Incomplete oxygen reduction and topotactic hydrogen~\cite{Si2019,Malyi2021} are theoretically expected to be present  because this is energetically favored,  at least for low Sr-doping. 
          {This leads to} {two holes in two orbitals forming a high-spin state}
          {and a three dimensional electronic structure, thus obstructing} {the intrinsic physics of superconducting nickelates.}


\section*{Conflict of Interest Statement}
The authors declare that the research was conducted in the absence of any commercial or financial relationships that could be construed as a potential conflict of interest.

\section*{Author Contributions}
All authors contributed to writing the article.

\section*{Funding}
We  acknowledge funding through the Austrian Science Funds (FWF) project numbers P  32044, {P 30213}, {Grant-in-Aids for Scientific Research (JSPS KAKENHI) grant number 19H05825}, JP20K22342 and JP21K13887. 
{OJ was supported by the Leibniz Association through the Leibniz Competition}.
{Calculations were partially
performed on the Vienna Scientific Cluster (VSC).}

\section*{Acknowledgments}
We thank {Atsushi Hariki, Motoaki Hirayama,} Josef Kaufmann,  Yusuke Nomura, and Terumasa Tadano for valuable discussions.

\section*{Data Availability Statement}
The present paper reviews previous work.  Datasets are accessible through the original publications.

\end{document}